# Control of Emerging-Market Target, Abnormal Stock Return: Evidence in Vietnam

*Correspondent author: Q. H. Van,*
*Ho Chi Minh City University of Technology and Education, Vietnam*
*Email quyenvh@hcmute.edu.vn*
*V. N. K. Tran, FPT University, Vietnam*

## Abstract

Joining with the upward trend of Global Foreign direct investment and FDI in emerging economies and emerging Asian economies, FDI to Vietnam, especially M&As have increased significantly in both numbers and value of deals from 1995 to 2015. Comparing with the deal triggered by a firm from Vietnam or another emerging country, the deal triggered by developed-market acquirer dominates the peers in term of return for acquirer's shareholder. In median, the acquirer's shareholder gains 21 cents every 1 US dollar spent to Vietnamese acquirer. Especially, the transfer in control of Vietnamese firm to developed-market acquirer drives the statistically significant return using the event study with OLS market model approach. The abnormal returns are 0.670%, 1.025%, 1.150% for two days, three days, and four days windows respectively.

# Abbreviations

**CAR – C**umulative **A**bnormal **R**eturn

**EEP - E**xtensive **E**conomic **P**rogram

**FDI – F**oreign **D**irect **I**nvestment

**GDP – G**ross **D**omestic **P**roduct

**GNP – G**ross **N**ational **P**roduct

**M&A – M**ergers and **A**cquisitions

**OLS – O**rdinary **L**east **S**quares

**SIC - S**tandard **I**ndustrial **C**lassification

**WTO** - **W**orld **T**rade **O**rganization

# 1. Introduction

In the late 1980s and early 1990s, many barriers, which prevent the capital flows especially foreign direct investment (FDI) from entering the emerging markets, were removed. The taking control activities from foreign country to emerging markets are also liberalized. Following these reforms, the FDI flows to emerging markets were expanded, especially in the form of foreign acquisitions (Chari et al., 2010). The foreign acquisition helps firm to extend its business across national border. With the emerging-market target, this extension gives acquirer a chance to run their business in a new market, where the differences in institutions and property rights protection are significant. By utilizing their advantages in management and technology, acquirers and their subsidiaries (formerly was the independent targets) could create more value than two separate entities before. These gains should be reflected in acquirers' stock price.

The motivation of this paper for considering M&A deals in Vietnam is that Vietnam is one of the prospects for FDI in either M&A deals or Greenfield projects (United Nations, 2016) with an upward trend in both number and value of M&A deals from 1995 to 2015 (see Figure 4). It could be explained by the fact that Vietnam is a emerging country and also located in Southeast Asia. Recently, these two characteristics have been promising targets for M&A deals (see figure 2 and 3) (United Nations, 2007, 2008, 2009, 2010, 2012, 2013, 2015).

This paper aims to identify the effect of acquisitions, which are triggered by developed-market acquirers in emerging market, to the return of acquirers' shareholders. The Vietnam's market, which is also the emerging market, is chosen as a representative.

We will investigate the gain in the financial market of acquirer' shareholders in order to reach the main objective of this study. The methodology adopted in this paper is the event study with OLS market model in combing with multivariate regression.

The main research question is "Whether the acquisitions in Vietnam with control bring the positive returns to developed-market acquirers' shareholders?" The null hypothesis is "The acquisitions in Vietnam with control bring the positive returns to developed-market acquirers' shareholders".

## 2. Vietnam background

Vietnam is an emerging country located in Indochina Peninsula, Southeast Asia. It is about 330,000 square kilometers and has a population of over 91 million (World Bank, 2016). After gaining independence in 1975, Vietnam rebuilt itself from devastation in every aspect from economy to politics and infrastructure. Since 1976, Vietnam has been officially named as the Socialist Republic of Vietnam. At this time, Vietnam had a centrally planned economy - a popular model of the economy in East Europe - until the extensive reform program in 1986 (also known as "Doi Moi"). Since "Doi Moi" started, Vietnam has embarked on building economic infrastructure such as the new regulation framework for a smoother operation of the market economy.

### 2.1. "Doi Moi" – Vietnam's Extensive Economic Program

The analysis of disadvantages of the central economy from 1975 to 1986 is shown in the work of Vuong (2004). This struggling period is characterized by researchers with economic inefficiencies, bureaucratism, overwhelming institutional rigidity, lack of functional market and also market-pricing system. In addition, private property rights such as possession of productive physical assets were not formally mentioned in laws and accepted by regulations. At that time, Vietnam was relying on Soviet heavily for financial assistances and aids. Moreover, Kimura (1986) also highlight the inefficiencies of Vietnam economy by the zero growth rate from 1975 to 1980 when reviewing the period from 1975 to 1986.

Vietnam gained the brighter picture when the **Extensive Economic Program** (EEP) was launched in 1986. At this point of time, a shift from backward market to a market economy has already been the target of political leaders. This direction was also advocated by major economic researchers. Vietnam has embarked on building economic infrastructure such as new regulation framework for a smoother operation of the market economy. At first, the laws on foreign investment in Vietnam was issued in 1987 by National Assembly. The amendments to this law was issued in the next following years (in 1990, 1992, 1996, 2000, and 2005). Moreover, the Constitution, which is also known as the foundation and the highest level of Law in Vietnam, was amended in 1992 with many important aspects reflecting the change in political point of view. The allowance for foreign investment is a clear evidence for that important amendment. The traits of centrally planned economy were blurred. The government recognize the legitimate rights of private properties and accept private economic sector. Figure 1 provides evidence for the success of the EEP. Except for the transitory period in the beginning, from 1989, the increasing trend of Vietnam GDP per capita keeps stable until now.

**Figure 1: Vietnam's GDP per capita in current U.S. dollar from 1985 to 2015**

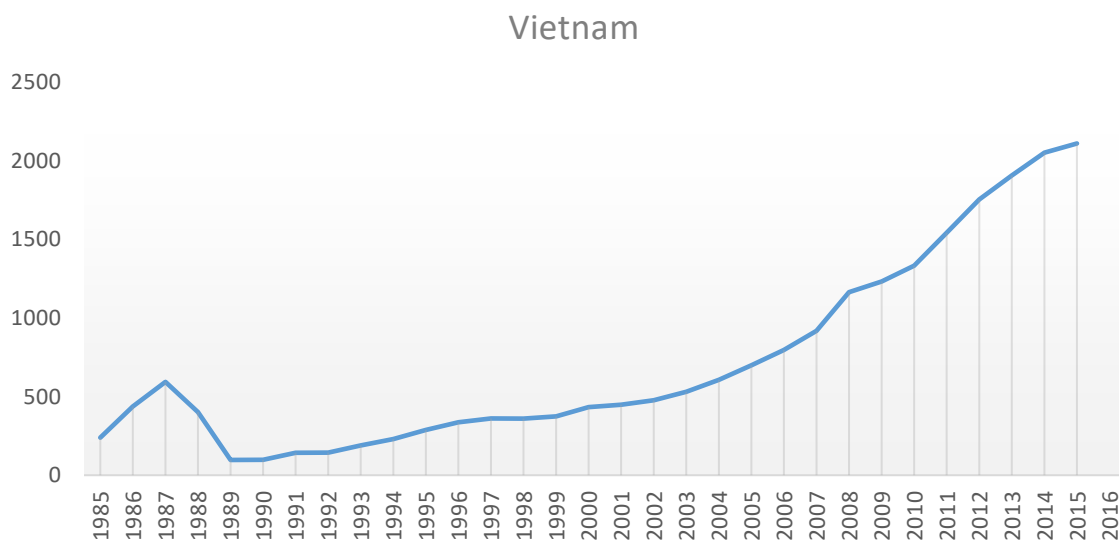

*Source: World Bank (2016)*

### 2.2. Securities market and WTO, two factors boosted Vietnam economy

Vietnam securities market was established in 2000, which is another important infrastructure for investing activities. The Decree 1998 issued by The Government about securities and securities market included more regulation for transferring ownership between entities in Vietnam capital market. This Decree also regulated the investing activities for foreign investors including individuals, funds, and organizations.

An important milestone for Vietnam economy in 2007 is when Vietnam became a 150$^{th}$ member of World Trade organization (also known as WTO) after 11 years of preparation. To be qualified as a WTO member, Vietnam has committed conditions with other member countries as well as WTO. For instance, Vietnam has committed to treat, offer the opportunity to every WTO member country equally, to reduce the tariffs, duties, and so on. Because of better economic infrastructure such as investment law, investor protection, and commitments when participating in WTO, especially for the foreign sector, foreign investors became more willing to invest in Vietnam. Consequently, FDI to Vietnam experienced a dramatic increase. In Figure 4, the value of M&A transaction in Vietnam in 2007 was 3 times higher than that in 2006 and more than total value of transaction from 1995 to 2006. The increasing trend has been continued for the later period

**Figure 2: FDI inflows, global and by group of economies, 1995−2015 (Billions of dollars)**

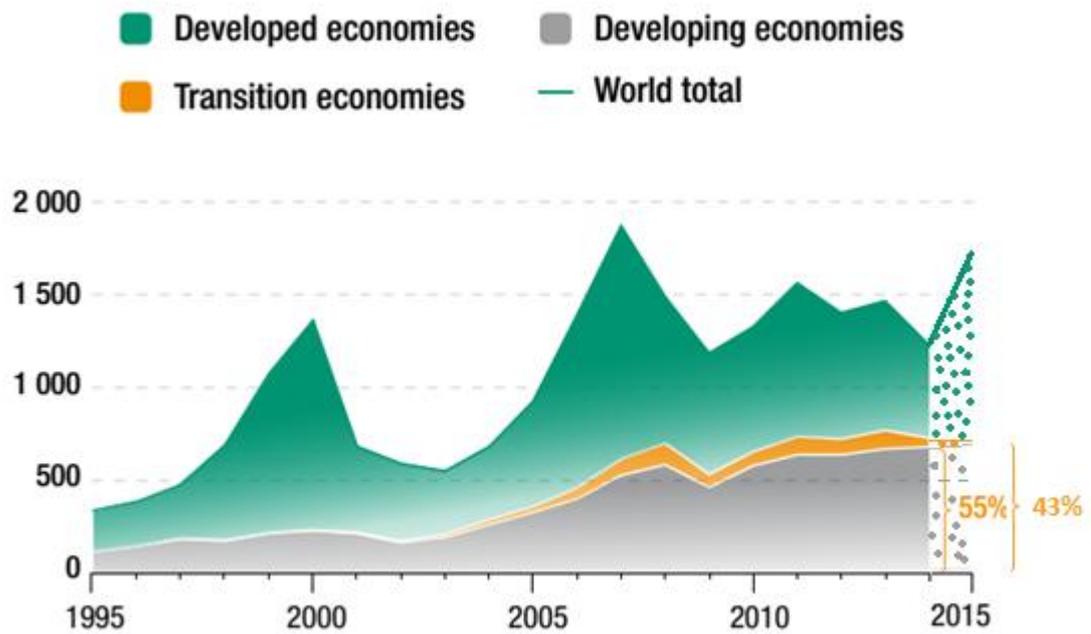

*Source:* UNCTAD *(United Nations, 2015b, 2016)*

**Figure 3: FDI inflows, by region, 2013–2015 (Billions of dollars)**

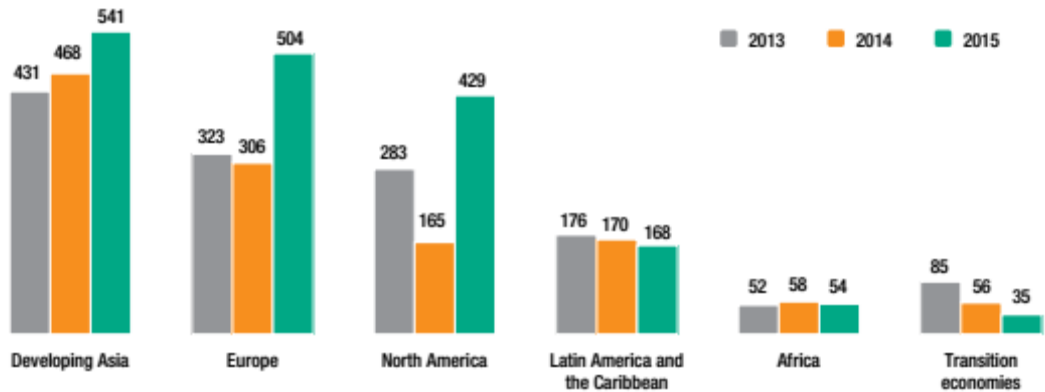

*Source: UNCTAD (United Nations, 2016)*

**Figure 4: Vietnam's M&A activities from 1995 - 2015**

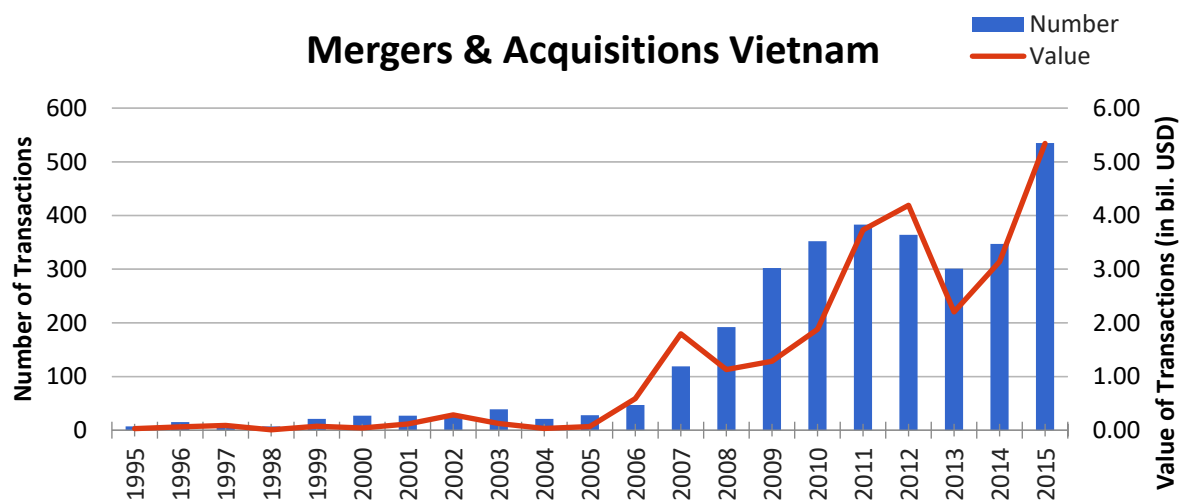

*Source: Institute for Mergers, Acquisitions and Alliances (IMAA 2017)*

### 3. Related literature review

#### 3.1. Mergers and Acquisitions, whether firm should do the cross-border deal?

Theoretically, previous researchers have provided many possible economic-based reasons why firms do M&As such as in order to increase their market power. Forming monopolies or oligopolies may also be one of their purposes. Firm also enrolls in M&A in order to improve their efficiency (Renneboog and Vansteenkiste, 2019). Depending on which type of deal (e.g. horizontal, vertical, or conglomerate), they could benefit in some different aspects (e.g. economies of scale and scope, learning economies or transaction cost). Taking advantage of diversification or managing risk is also the possible reason (Andrade et al., 2001).

Moreover, previous researchers question whether firm does M&A with domestic or foreign target. In theory, acquirer chooses the best option which could bring the highest net present value. Acquirer considers acquiring a foreign target when the other options in domestic market do not offer them higher net present value. Although the benefit for acquirer seems to be clear, in fact, it is much more complicated. When doing deal in the foreign markets, bidders have to challenge themselves to be familiar with the new environments. Besides, bidders, who are strangers to the market, also face the informative asymmetry so that it is possible for them to pay a higher price in order to acquire the target.

### 3.2. Winners and Losers in the Merger Game

#### 3.2.1. The whole picture

Martynova and Renneboog (2008) review a large number of previous researches covering five major waves of M&A. They find that the bidders' benefits, which are indicated by short-term signal (e.g. share-price-abnormal return) and long-term signal (e.g. accounting indicators), depend on the wave that the deal belongs to and the method to measure. Also, some researchers point out the under- or over-performance comes from specific factors (e.g. acquirers' size (Moeller et al., 2004) or payment methods (Dutta and Jog, 2009)).

However, the viewpoint that M&A could bring advantages to targets are popularly agreed by many previous researchers (Renneboog and Vansteenkiste, 2019). In the review of Martynova and Renneboog (2008), there is a negligible disadvantage for the M&A's target. Out of 38 journals having results for short-term target return, 36 journals prove the statistically significant positive results. There are only two journals giving mixed results. In addition, when investigating 1273 M&A transactions triggered by the US firms from 1970 to 1987, Harris and Ravenscraft (1991) find the positive return of target is boosted if the targets from foreign countries.

#### 3.2.2. The win-win game for developed-market acquirer and emerging market target

Considering a smaller geographic scope of M&As, the transaction between developed-market acquirer and emerging-market target offers a brighter scenario for the acquirer. Chari et al. (2010) find the positive return for developed-market acquirer when taking control emerging-market target from 1985 to 2005. Comparing the developed-emerging market acquisition with developed-developed and emerging-emerging market pairs, Chari et al. (2010) finds that positive gain is unique to the developed-emerging market acquisition. The increased value may be created by improving the target value through sharing acquirer's better institutional and corporate governance practices. The value created is also boosted if the difference in intangible asset level is significant. In addition, the positive return for target stays consistent. Lebedev et al. (2015) review 51 journals that focus on M&As in emerging countries and point out that the return for the target is typically positive.

### 3.3. Factors affect transaction's return

**Controlling right:** Acquirer tends to help the target unconditionally just when they have a close-knit relationship (i.e. having the controlling right). By utilizing - their own advantages in management and technology, acquirers from developed markets and their subsidiaries (formerly was the independent target) could create more value than two separate entities before. We expect the positive relationship between the controlling right and acquirer's return here.

**The private target:** Private firms cannot be bought and sold as easily as public firms. The acquirer gets a discount for this lack of liquidity. (Fuller et al., 2002; Chen et al., 2019). It is generally accepted that announcement returns for acquiring large and public targets are normally negative, and conversely positive when acquiring small and private targets (Schneider and Spalt, 2017). We also expect the positive relationship between the private target and acquirer's return here.

**Diversifying deal:** Doing main-business deal - could bring to acquirer more value than diversifying deal as main-business deal increases the size and breadth of firm and consequently, it will receive the benefit from economies of scale (Singh and Montgomery, 1987). We expect the negative relationship between the diversifying deal and acquirer's return.

**Acquirer's size:** The abnormal return associated with acquisition for small firm dominates that for large firms (Li et al., 2018). Large firms offer larger acquisition premiums than small ones and enter deals with negative dollar synergy gains. Managerial hubris plays vital role in the decisions of large firms (Moeller et al., 2004). We expect the negative relationship between the acquirer's size and its return.

**Acquirer's industry:** In highly concentrated industry, M&A could bring higher market power to acquirers because of the collusion theory (Roller et al., 2001). In converse, concentrated industry could reduces the firms' incentive to make an M&A deal for improving their performance (Porter, 1990). Also, the relationship between the acquirer and target causes the difference in the deal's performance.

### 3.4. Methodology and limitations

In M&A literature relating to investigate the gain of acquirer and target, there are two main approaches: stock-return (e.g. Alexandridis et al., 2017) and accounting-based method (e.g. Malmendier et al., 2018). The method based on stock-return requires that the considering firms must be public companies. However, based on the efficient market theory (Fama, 1970), the market reflexes the fundamental factors efficiently just when the market is in strong form. The truth is that the number of the stock markets in the world, which are in strong form, is rare, so that the use of this method is biased and does not give the sufficient evaluation. There are some researches that suggest using accounting-based methods such as productivity, profitability or innovation indicators. Because these methods could reflect a better picture. However, they are still ambiguous because sometimes using two different accounting-based methods lead to different results (Martynova and Renneboog, 2008). Also, there are some complexities with accounting-based method. For example, when a bidder does more than one deal including a deal to emerging Asian market, it is difficult to isolate the effect of the deal to emerging Asia from the other deals.

## 4. Data and research methodology

### 4.1. Data source

Following Netter et al. (2011), the data of M&A deals was collected from Thomson One Banker (TOB) database of Reuters. We collect daily trading data (excluding Vietnam) from Datastream of Reuters. For Vietnam's trading data, I collect from the Vietnam's stock exchanges. The market benchmark (index) for each acquirer or target is the broadest index of the stock exchange in the acquirer's or target's country. We exclude the trading data of the non-working days (e.g. Lunar new year in some Asian countries or Bank holiday in the UK).

There are three samples of considering M&A deals, which consist of publicly listed acquirers and any Vietnamese targets (public or private) announced between 1995 and 2015. Sample 1 (DM-VN) is the main sample of this study in which every considering deal is triggered by a developed-market acquirer. Sample 2 (EM-VN) includes deals triggered by emerging-market acquirers (excluding Vietnam). Sample 3 (VN-VN) includes deals triggered by Vietnamese firms.

The developed-market nations in the samples are Australia, Canada, Denmark, France, Germany, Hong Kong, Japan, Korea, Netherlands, Singapore, Sweden, Taiwan, The US and The UK. The emerging-market nations in the samples include India, Indonesia, Malaysia, Philippines, Russia, and Thailand (see Table 2). The criterion for classifying country into each above group is based on the World Bank Analytical Classifications (2016). The country is classified as developed country when its GNP per capita is considered as high. Table 11 and Table 12 in Appendix give further detail about classification criteria as well as the country's classification.

To be included in any of the three above samples, the observations must meet these following criterions:1) the target must be from Vietnam, 2) the deal was announced between 1995 and 2015 and must be successful in this time period, 3) the acquirer must be publicly listed, 4) the acquirer must not have the control in the target before the transaction, 5) there are at least 195 trading days (equivalently nine months) before the announcement day for estimating the intercept and slope of the market model.

**Table 1: Number of transactions by nation in three samples**

|  | **Sample 1 (DM-VN)** | **Sample 2 (EM-VN)** | **Sample 3 (VN-VN)** |
| --- | --- | --- | --- |
| Sample description | Developed-market acquirers and Vietnamese targets | Emerging-market acquirers and Vietnamese targets | Vietnamese acquirers and Vietnamese targets |
| **Acquirer nation** | Australia (11), Canada (2), Denmark (1), France (13), Germany (2), Hong Kong (1), Japan (32), Korea (9), Netherlands (2), Singapore (12), Sweden (1), Taiwan (6), The US (12) and The UK (8). | India (1), Indonesia (1), Malaysia (10), Philippines (3), Rusia (2) and Thailand (5). | Vietnam (174) |
| **Target nation** | Vietnam (112) | Vietnam (22) | Vietnam (174) |

*The table presents the acquirer's origins of three samples of transactions involving publicly listed acquirers and any Vietnamese targets (public or private) announced between 1995 and 2015. Sample 1 (DM-VN) is the main sample in which every considering deal is triggered by a developed-market acquirer. Sample 2 (EM-VN) considers deals triggered by an emerging-market acquirer(excluding Vietnam). Sample 3 (VN-VN) includes deals triggered by Vietnamese firms. The numbers in parentheses after nations' name are the number of deals from that countries.*

In general, there are 4443 M&A transactions involving Vietnamese target in TOB database. Applying the next two aforementioned criterions that are deal was announced in the period of 1995 - 2015 and must be successful in this period of time, there are 3853 and 2370 transactions respectively. As for the third criterion, there are 762 deals triggered by public acquirers. In these 762 deals, there are 655 deals that the acquired share after the transaction is at least 5%. The percentage of 5% is chosen as sorting criteria because a shareholder or a group of shareholders holding at least 5% could have some special rights in the target (e.g. convene an unusual-shareholder meeting, nominate a person to board of directors). After excluding the transactions in which the acquirers already have had the control before the transactions, there are 583 deals for the three samples. The last criterion is that there is no other event that could affect the acquirer's stock price in 15 days before and 15 days after the announcement day of the M&A transaction. There are finally 112, 22 and 174 deals come to the sample DM-VN, EM-VN, VN-VN respectively. These steps above are summarized in Table 3 below.

**Table 2: Criteria applied and changing in observations**

| Criteria | | Number of M&A transaction | |
|---|---|---|---|
| Vietnamese target | | 4443 | |
| Announced between 01/01/1995-12/31/2015 | | 3853 | |
| Affected between 01/01/1995-12/31/2015 | | 2370 | |
| Public acquirer | | 762 | |
| The acquired share after transaction is at least 5% | | 655 | |
| Acquirer does not have control in target before transaction | | 583 | |
| No other event | | 308 | |
| | **Sample 1 (DM-VN)** | **Sample 2 (EM-VN)** | **Sample 3 (VN-VN)** |
| Sample description | Developed-market acquirers and Vietnamese targets | Emerging-market acquirers and Vietnamese targets | Vietnamese acquirers and Vietnamese targets |
| | 112 | 22 | 174 |

*Source: Thomson one banker, Reuters (2017).*

### 4.2. The first evidences of gains for developed-market acquirers

Table 3 summarizes statistics for three samples of M&A deals involving publicly listed acquirers and any Vietnamese targets (public or private), announced between 1995 and 2015. Comparing the median numbers, the transaction size of developed-market acquirers is approximately four times higher than that of emerging-market acquirers (excluding Vietnamese firm) and the transaction size of emerging-market acquirers is approximately four times higher than that of Vietnamese acquirers. The market capitalization for developed-market acquirers are also far higher than their peers, particularly, 10 and 80 times higher than emerging-market and Vietnamese acquirers respectively.

The proportions of private targets are 87.5% for developed-market acquirers and 95.5% emerging-market acquirers while this proportion for the Vietnamese acquirers is 62% (Table 4). Clearly, the proportion of private target is significantly higher for foreign acquirer than Vietnamese acquirer. The reason is that there was a restriction for cross-border acquirers to take a control of public targets in Vietnam. Vietnam's law on foreign investment did not allow the foreign investors including individuals, institutions, and funds, etc. hold more than 50% of Vietnam's public firm prior to September 2015 (there is a foreign holding cap at 50%). From September 2015 onward, this cap has been removed and the foreign acquirers could hold up to 100 percent (see Vietnam, The Government, 2015). However, this removal needs more time to be completely effective.

Table 3 also suggests that the Vietnamese acquirers are more likely to do deals in order to diversify their business, while other emerging-market acquirers are least likely. Deals are considered as diversifying acquisition if the acquirer and target are in the different three-digit SIC code.

Table 3 also shows the distribution of transactions by industrial sectors of acquirers and targets. In general, each group of considered acquirers had different taste of interests. Financial services attracts the most interests from the developed-country acquirers. More than 29% transactions triggered by developed-market acquirers and also 29% of the targets bought by developed-market acquirers are operating in the financial services sector. Basic manufacturing attracts the most interests from emerging-market acquirers. More than 27% transactions triggered by emerging-market acquirers and also over 31% target bought by developed-market acquirers are operating in the basic manufacturing sector. Vietnamese acquirers were most likely to operate in Agriculture and consumer products. More than 27% transactions triggered by Vietnamese acquirers and also over 26% target bought by Vietnamese acquirers are operating in agriculture and consumer products sector.

Considering the taking-control transaction, the positive median of all three windows CAR is unique for developed-market acquirer. Particularly, the developed-market acquirers get the positive value of 0.670%, 1.025%, and 1.150% for the three windows comparing with the negative value of -0.060%, -0.215% and -0.520% for emerging-market respectively. These CAR value for Vietnamese acquirers are -0.165%, -0.310 and -0.105%. In addition, these CARs values are statistically significantly lower for Vietnamese acquirers comparing with developed-market acquirers and also for emerging-market acquirers comparing with developed-market acquirers for all the three windows. However, the difference between Vietnamese acquirers and emerging-market acquirers is not statistically significant.

**Table 3: Summary statistics: Firm and deal characteristics**

|  | Sample 1 (DM-VN) | Sample 2 (EM-VN) | Sample 3 (VN-VN) |
|---|---|---|---|
| Sample description | Developed-market acquirers and Vietnamese targets | Emerging-market acquirers and Vietnamese targets | Vietnamese acquirers and Vietnamese targets |
| *Firm and deal characteristics* | | | |
| Median transaction size ($M) | 18.564 | 4.816 | 1.21 |
| Median acquirer market capitalization ($M) | 3764.355 | 329.690 | 40.255 |
| Private target (%) | 87.500% | 95.45% | 62.069% |
| Diversifying acquisition (%) | 52.679% | 27.273% | 71.264% |
| Median acquirer CAR (%) | | | |
| Window (0, 1) | 0.145% | -0.060% | -0.122% |
| Window (-1, 1) | 0.120% | 0.300% | -0.333% |
| Window (-2, 1) | 0.460% | 0.650% | -0.491% |
| Median acquirer CAR (%) with control acquired | | | |
| Window (0, 1) | 0.670% | -0.060% | -0.165% |
| Window (-1, 1) | 1.025% | -0.215% | -0.310% |
| Window (-2, 1) | 1.150% | -0.520% | -0.105% |
| *Acquirer industry* | | | |
| Agriculture and consumer products | 8.929% | 0.000% | 27.011% |
| Basic manufacturing | 18.750% | 27.273% | 17.241% |
| Machinery and electronics | 12.500% | 18.182% | 11.494% |
| Utilities and transportation | 7.143% | 4.545% | 10.345% |
| Wholesale and retail trade | 11.607% | 13.636% | 4.023% |
| Financial services | 29.464% | 22.727% | 24.138% |
| Tourism and miscellaneous services | 11.607% | 13.636% | 5.747% |
| *Target industry* | | | |
| Agriculture and consumer products | 8.929% | 9.091% | 26.437% |
| Basic manufacturing | 20.536% | 31.818% | 16.667% |
| Machinery and electronics | 12.500% | 18.182% | 12.069% |
| Utilities and transportation | 8.036% | 4.545% | 13.793% |
| Wholesale and retail trade | 6.250% | 4.545% | 10.345% |
| Financial services | 29.464% | 22.727% | 14.943% |
| Tourism and miscellaneous services | 14.286% | 9.091% | 5.747% |

*The table shows three samples of deals involving publicly listed acquirers and any Vietnamese targets (public or private), announced between 1995 and 2015. CARs are estimated using a two-day (0,1), three-day (-1,1) and four-day (-2, 1) event window. The diversifying acquisition is a transaction that acquirer and target are in the different three-digit SIC industry Code. The SIC is divided into 7 groups (as Chari et al., 2010): Agriculture and Consumer products include firms with SIC code start by 00–19; basic manufacturing starts by 20–29; machinery and electronics start by 30–39; utilities and transportation start by 40–49; wholesale and retail trade start by 50–59; Finance, Insurance and Real Estate (FIRE) start by 60–69; tourism and miscellaneous services start by 70–99. The proportion of each industry, Private target, and Diversifying acquisition are calculated based on the number of transactions.*

The table 4 shows more detail about CARs (mean) of the general economy and every sector of three samples of deals involving publicly listed acquirers and any Vietnamese targets (public or private), announced between 1995 and 2015. The taking control transactions are separated in the column named Control. The t-test with one tail test is adopted to denote the statistical significance of the results. Considering the performance of the developed-market acquirers (sample 1), who do taking-control deals, although the whole sample shows the highly significant benefit, the gain in each sector is mixed. Some sectors have positively significant CARs in all three windows: basic manufacturing and utilities and transportation. Some sectors just have significantly positive CARs in two or one windows, such as Wholesale and retail trade or Agriculture and consumer products. Whereas, the reminders have positive values, even negative but are not significant. This complicated result may be supported by the difference in acquirers' industries, which have different intangible asset level.

Moreover, at the first glance through table 4, the domination is clear for developed-market bidders in term of shareholders' return aspect. There is just a small benefit for emerging-market acquirer in machinery and electronics sector while this minor gain for Vietnamese bidders is in financial services sector.

**Table 4: Summary statistics: CARs by sector for three samples**

|  | Windows | Sample 1 (DM-VN) Developed-market acquirers and Vietnamese targets | | Sample 2 (EM-VN) Emerging-market acquirers and Vietnamese targets | | Sample 3 (VN-VN) Vietnamese acquirers and Vietnamese targets | |
|---|---|---|---|---|---|---|---|
|  |  | All deal | Control | All deal | Control | All deal | Control |
| *(All)* | (0,1) | **0.002** | **0.013*** | -0.005 | -0.006 | -0.001 | -0.001 |
|  |  | **(0.003)** | **(0.004)** | (0.006) | (0.008) | (0.003) | (0.004) |
|  | (-1,1) | **0.006*** | **0.021*** | -0.012 | -0.018 | **-0.006**** | -0.004 |
|  |  | **(0.004)** | **(0.007)** | (0.012) | (0.018) | **(0.003)** | (0.004) |
|  | (-2,1) | **0.007**** | **0.020*** | -0.006 | -0.015 | **-0.006*** | -0.002 |
|  |  | **(0.004)** | **(0.007)** | (0.013) | (0.019) | **(0.004)** | (0.005) |
|  | N0 | 112 | 50 | 22 | 14 | 174 | 90 |
| *Acquirer industry* | | | | | | | |
| **Agriculture and consumer products** | (0,1) | **0.017**** | **0.018*** | - | - | -0.000 | 0.000 |
|  |  | **(0.008)** | **(0.010)** |  |  | (0.006) | (0.010) |
|  | (-1,1) | **0.037*** | 0.036 | - | **-** | **-0.010*** | -0.007 |
|  |  | **(0.022)** | (0.027) |  |  | **(0.007)** | (0.012) |
|  | (-2,1) | **0.033*** | 0.031 | - | **-** | **-0.011*** | -0.008 |
|  |  | **(0.024** | (0.030) |  |  | **(0.008)** | (0.012) |
|  | N1 | 10 | 8 | 0 | 0 | 47 | 20 |
| **Basic manufacturing** | (0,1) | **0.005*** | **0.012*** | **-0.024*** | **-0.024*** | -0.004 | **-0.011**** |
|  |  | **(0.004)** | **(0.004)** | **(0.013)** | **(0.013)** | (0.005) | **(0.006)** |
|  | (-1,1) | 0.004 | **0.010**** | -0.052 | -0.052 | 0.005 | **-0.011**** |
|  |  | (0.005) | **(0.004)** | (0.040) | (0.040) | (0.006) | **(0.006)** |
|  | (-2,1) | **0.007*** | **0.11*** | -0.053 | -0.053 | -0.004 | -0.003 |
|  |  | **(0.004)** | **(0.004)** | (0.040) | (0.040) | (0.007) | (0.008) |
| | N2 | 21 | 15 | 6 | 6 | 30 | 15 |
| **Machinery and electronics** | (0,1) | -0.008 (0.006) | -0.000 (0.005) | **0.011*** **(0.006)** | **0.014*** **(0.007)** | **-0.020**** **(0.007)** | **-0.017*** **(0.011)** |
| | (-1,1) | -0.005 (0.007) | 0.001 (0.007) | 0.004 (0.011) | **0.013*** **(0.007)** | **0.021**** **(0.010)** | -0.014 (0.013) |
| | (-2,1) | 0.002 (0.010) | 0.007 (0.008) | 0.002 (0.015) | 0.015 (0.011) | -0.012 (0.012) | 0.003 (0.017) |
| | N3 | 14 | 8 | 4 | 3 | 20 | 11 |
| **Utilities and transportation** | (0,1) | 0.008 (0.0017) | **0.035*** **(0.021)** | -0.008 (-) | - | 0.005 (0.008) | -0.002 (0.008) |
| | (-1,1) | 0.026 (0.015) | **0.056**** **(0.014)** | 0.003 (-) | - | -0.005 (0.010) | 0.001 (0.011) |
| | (-2,1) | 0.013 (0.011) | **0.031**** **(0.011)** | -0.002 (-) | - | -0.008 (0.011) | -0.006 (0.016) |
| | N4 | 8 | 4 | 1 | 0 | 18 | 8 |
| **Wholesale and retail trade** | (0,1) | -0.008 (0.010) | 0.012 (0.009) | -0.013 (0.009) | -0.013 (0.009) | 0.010 (0.020) | 0.011 (0.029) |
| | (-1,1) | -0.009 (0.013) | **0.026**** **(0.010)** | -0.006 (0.007) | -0.006 (0.007) | 0.013 (0.026) | 0.010 (0.038) |
| | (-2,1) | -0.002 (0.008) | **0.027*** **(0.014)** | -0.005 (0.010) | -0.005 (0.010) | 0.008 (0.025) | 0.006 (0.033) |
| | N5 | 13 | 4 | 3 | 3 | 7 | 5 |
| **Financial services** | (0,1) | -0.000 (0.003) | -0.009 (0.010) | 0.004 (0.006) | 0.014 (-) | 0.005 (0.005) | **0.009*** **(0.006)** |
| | (-1,1) | 0.002 (0.004) | 0.007 (0.020) | **0.011**** **(0.004)** | 0.005 (-) | 0.000 (0.006) | 0.003 (0.007) |
| | (-2,1) | 0.002 (0.06) | 0.019 (0.027) | **-0.021**** **(0.006)** | 0.011 (-) | -0.001 (0.008) | 0.003 (0.011) |
| | N6 | 33 | 4 | 5 | 1 | 42 | 25 |
| **Tourism and miscellaneous services** | (0,1) | 0.006 (0.014) | 0.027 (0.021) | 0.003 (0.024) | 0.044 (-) | -0.007 (0.005) | -0.002 (0.005) |
| | (-1,1) | 0.008 (0.020) | 0.032 (0.030) | 0.001 (0.025) | 0.038 (-) | -0.004 (0.011) | -0.005 (0.016) |
| | (-2,1) | 0.008 (0.022) | 0.032 (0.034) | 0.028 (0.018) | 0.065 (-) | -0.006 (0.012) | -0.007 (0.009) |
| | N7 | 13 | 7 | 3 | 1 | 10 | 6 |

*The table shows the general economy's and every sector's CARs (mean) in three samples (described in the header row) of deals involving publicly listed acquirers and any Vietnamese targets (public or private), announced between 1995 and 2015. The taking control transactions are separated in the column named as Control. CARs are estimated using (0,1), (-1,1) and (-2,1) windows around the announcement date. The SIC is divided into 7 groups (as Chari et al., 2010): Agriculture and Consumer products starts by 00–19; basic manufacturing starts by 20–29; machinery and electronics starts by 30–39; utilities and transportation starts by 40–49; wholesale and retail trade starts by 50–59; Finance, Insurance and Real Estate (FIRE) starts by 60–69; tourism and miscellaneous services starts by 70–99. Mean value of CARs are calculated. Standard errors are reported in parentheses. ∗, ∗∗ and ∗∗∗ denote statistically significant levels at 10%, 5%, and 1% respectively.*

Table 5 shows the change in pre- and post-acquisition ownership resulting from M&A transactions in the three samples. The columns and the rows indicate the pre- and post-acquisition ownership, respectively. Column two and three break down the pre-acquisition ownership into two categories. Particularly, in column two, the acquirer did not have any ownership staking in the target before the transaction. Column three shows the numbers for transactions where the acquirer did have the minor ownership in the target before the announcement. Column 4, 5 and 6 break down the number in column 3 into three smaller groups of pre-acquisition minor ownership.

At first glance, the number of transactions which acquirers already have had the ownership before is significant, which accounts for about one-fourth of total transactions in three samples. This support the evidence that the acquirers have pre-acquisition experience before the M&A deals

**Table 5: Pre- and post-acquisition ownership by three samples**

| Number of M&A transactions | Acquirer had minority interest before acquisition | | Pre-acquisition ownership | | |
|---|---|---|---|---|---|
| Post-acquisition ownership | No | Yes | <20% | 20–40% | 40–50% |
| **Panel A: Sample 1 (DM-VN)** | | | | | |
| **0-50%** | 53 | 9 | 8 | 1 | 0 |
| **50–95%** | 21 | 5 | 1 | 2 | 2 |
| **95–100%** | 22 | 2 | 0 | 1 | 1 |
| **Panel B: Sample 2 (EM-VN)** | | | | | |
| **0-50%** | 8 | 0 | 0 | 0 | 0 |
| **50–95%** | 3 | 2 | 0 | 1 | 1 |
| **95–100%** | 9 | 0 | 0 | 0 | 0 |
| **Panel C: Sample 3 (VN-VN)** | | | | | |
| **0-50%** | 59 | 25 | 17 | 8 | 0 |
| **50–95%** | 25 | 30 | 0 | 17 | 13 |
| **95–100%** | 30 | 5 | 0 | 3 | 2 |

*The table classifies M&A transactions by three samples and by pre- and post-acquisition ownership. This table covers all M&A transactions involving publicly listed acquirers and any Vietnamese targets (public or private), announced between 1995 and 2015 and for which ownership data are available. Control information is the items "percent shares acquired in transaction" and "percent shares owned after transaction" collected from Thomson One Banker database*

### 4.3. Event study

#### 4.3.1. Methodology applied

##### 4.3.1.1. Event study with OLS market model

This research examines the abnormal return of M&A transaction using Market model (see Brown and Warner, 1980; Brown and Warner 1985) through event study. The first step is determining the windows. This research focuses on 3 groups of windows with their assumptions. The first group includes short-length windows around the announcement day (e.g. two-day (day [0] – [1]), three-day (day [-1] – [1]), four-day (day [-2] – [1]). This group is employed to test for the liquidity assumption of the acquirer's market. The second group includes windows, which start before the announcement day and end on day 0, in order to test whether there is a leaking information effect. They are two-day (day [-1] – [0]), three-day (day [-2] – [0]), six-day (day [-5] – [0]), ten-day (day [-10] – [-1]), eleven-day ([-10] – [0]) windows. The last group includes longer windows in order to test for the less liquidity of the market or to capture the thin trading (e.g. seven-day (day [-1] – [5]), twelve-day (day [-1] – [10]), eighteen-day (day [2] – [15]), twenty-one-day ([-10] – [10]). The estimation period is from month 9th (day [-196]) to month 3rd (day [-65]) before the announcement day (day [0]). This period is chosen in order to mitigate the bias problem caused by the M&A transactions to the alpha and beta from. The share price return in day [t] is calculated by dividing the logarithmic ratio of the price on day [t] by price on day [t-1]. The abnormal return is the difference between the actual return and the predicted return estimated by the market model (see Brown and Warner, 1980; Brown and Warner 1985). The market index used in the market model (1) for each firm's estimated return is the broadest index of the stock market where that firm is listed.

$$AR_{it} = R_{it} - (\hat{a}_i + \hat{b}_i * R_{mt}) \qquad (1)$$

Where $AR_{it}$ is the abnormal return of firm i at time t, $R_{it}$ is the actual return of firm i at time t. $\hat{a}_i$ and $\hat{b}_i$ are the estimated intercept and slope of the OLS market model, respectively.

We use the t-test statistic to examine whether the cumulative average abnormal returns are statistically significant. For example, by applying the t test for one-day window (day [0]), we have:

$$\frac{\frac{1}{N}\sum_1^N AR_{i0}}{\frac{1}{N}(\sum_1^N[\frac{1}{130}\sum_{-196}^{-65}(AR_{it} - (\frac{1}{131}\sum_{-196}^{-65}(AR_{it}))^2])^{\frac{1}{2}}}$$

The t-test is appropriate because the distribution of all observations in this research is large and nearly symmetric. Table 10 supplies further information about the distribution such as kurtosis and skewness.

Combining descriptive statistics with modeling the abnormal return by multivariate regression (see Binder, 1985), this research aims to identify the benefits in the share price of acquirers. This research also aims to investigate the effects of every single variable on the performance of acquirers. There are some previous works conducting the similar tasks (e.g. Morck and Yeung, 1992; Moeller and Schlingemann, 2005; Du and Boateng, 2015).

### 4.3.1.2. Multi-variate regression model:

Binder (1985) suggests the detailed procedure of applying multi-variate regression model in event study. A number of hypotheses will be tested by this approach.

The example equation below stimulates the multi-variate regression model used in Table 9.

$$CAR_i = \alpha + \beta_1 control_i + \beta_2 DM\ acquirer_i + \beta_3 control_i * DM\ Acquirer_i + \beta_4 Listed_i + \beta_5 non-diversified_i + \beta_6 MV_i + \beta_7 MV * Control_i + \varepsilon$$

Where $CAR_i$ is the cumulative abnormal return of firm i in determined windows such as three or five-day event windows.

$Control_i$ is a dummy variable that takes the value of 1 if the acquirer holds at least 50% target's shares after the transaction, and 0 otherwise.

$DM\ acquirer_i$ is a dummy variable that takes the value of 1 if the acquirer is from the developed-market, and 0 otherwise.

$Control_i * DM\ Acquirer_i$ is an interactive term between 2 variables $Control_i$ and $DM\ acquirer_i$.

$Listed_i$ is a dummy variable that takes the value of 1 if the target is publicly listed in Vietnam stock exchange.

$Non-diversified_i$ is a dummy variable that takes the value of 1 if the acquirer i and target from the same industry (defined as in the same three-digit SIC industry code), and 0 otherwise.

$MV_i$ is acquirer's market capitalizations. This data is taken from DataStream of Reuters and is log-transformed.

$MV * Control_i$ is an interactive term between 2 variables $MV_i$ and $Control_i$.

The hypotheses constructed in this research follows the procedure of Chari et al. (2010).

**4.3.1.3. Validations of data for testing hypotheses**

To ensure the results of this research are accurate and significant, the conditions of adopted tests must be valid. For instance, to use T-test, the number of observations should be large. Another example is that the OLS market model for event study is ideally when the sample size is about 50. However, the smaller samples ranging from 5 to 20 are acceptable with the appropriate probability of Type I error (see Brown and Warner, 1985).

**Samples' distribution**

Table 10 (see appendix) shows some statistics for the distributions of three samples.

The whole sample has 308 observations. The shape of distributions is relatively symmetric for all considering windows. With all the kurtoses are higher than 3 (a characteristic of normal distribution according to Aczel and Sounderpandian (2008)), the distributions are considered as leptokurtic.

The DM-VN sample has 112 observations. The shape of distributions is relatively right-skewed for all considered windows. With all the kurtoses are higher than 3, the distributions are leptokurtic.

The EM-VN sample has 22 observations. The shape of distributions is relatively left-skewed for all considered windows. With all the kurtoses are higher than 3, the distributions are leptokurtic.

The VN-VN sample has 174 observations. The shape of distributions is relatively symmetric for all considered windows. With all the kurtoses are higher than 3, the distributions are leptokurtic.

Using "sktest" command for normality test in Stata (see D'Agostino et al., 1990), all the 48 tests applying on 12 windows in 4 sample groups can reject the normally distributed hypothesis. However, this is not unexpected.

The fact is that with the small sample size, the sample mean excess return is not normally distributed. However, when the number of securities increases, this distribution converges to normal distribution. The mechanism is explained using The Central Limit Theorem (see Billingsley, 1979). Considering cross-section of securities, if the excess returns are independent and their distributions are identically drawn from finite variance distributions, the sample mean excess return's distribution converges to normal distribution as the number of securities increases (see Brown and Warner, 1985). Hagerman (1978) also supports this convergence by using a sample of 286 securities traded on the American Stock Exchange and 805 securities traded on the NYSE, from 1962 to 1976.

**Simple correlation analysis for Multi-variate regression model**

Ideally, the collinearity in Multi-variate regression model should not exist. The collinearity is an interaction between independent variables in the model. If this interaction is large and significant, the model may be not significant and meaningless (see Aczel and Sounderpandian, 2008). As for mitigating the collinearity, one of the best ways is dropping collinear variables from the regression equation. Aczel and Sounderpandian (2008) also discuss more methods to overcome this situation.

Table 6 shows the high correlation between $Control_i$ and $MV * Control_i$ (0.810), $DM\ acquirer_i$ and $MV_{ij}$ (0.649), $MV * Control_i$ and $Control_i * DM\ Acquirer_i$ (0.639), $DM\ acquirer_i$ and $Control_i * DM\ Acquirer_i$ (0.582), $Control_i * DM\ Acquirer_i$ and $Control_i$ (0.440) due to the effect of dummy variables and interactive firms between two variables. However, the collinearity, which may cause the model to be meaningless, does not exist. Using these dummy variables is necessary for answering the research question (see Chari et al., 2010). Moreover, some methods of mitigating the collinearity are adopted and the conclusions, which clarify the strength or meaningfullness of the models, before and after conducting these methods are similar

**Table 6: Correlation matrix for Multi-variate regression model**

|  | $Control_i$ | $DM\ acquirer_i$ | $Control_i * DM\ Acquirer_i$ | $Listed_i$ | $Non-diversified_i$ | $MV_{ij}$ | $MV * Control_i$ |
|---|---|---|---|---|---|---|---|
| $Control_i$ | 1 | | | | | | |
| $DM\ acquirer_i$ | -0.081 | 1 | | | | | |
| $Control_i * DM\ Acquirer_i$ | 0.440 | 0.582 | 1 | | | | |
| $Listed_i$ | -0.199 | -0.237 | -0.263 | 1 | | | |
| $Non-diversified_i$ | 0.193 | 0.135 | 0.121 | -0.171 | 1 | | |
| $MV_{ij}$ | -0.124 | 0.649 | 0.248 | -0.227 | 0.108 | 1 | |

| | | | | | | | | |
|---|---|---|---|---|---|---|---|---|
| $MV * Control_i$ | 0.810 | 0.151 | 0.639 | -0.250 | 0.174 | 0.267 | 1 | |

## 5. Results

### 5.1. Returns for developed-market acquirers in emerging-market acquisitions

Table 4 shows the only positive CARs by median is for the developed-market acquirers. In addition, Table 5 indicates the clear domination for developed-market bidders in term of shareholders' return aspect. These results suggest the unique of benefit for developed-market acquirers in taking acquisitions with Vietnamese firms. Table 7 will give further information for M&A transaction triggered by developed-market acquirers to Vietnam. The series of regressions with CAR employed as the dependent variable are run for three windows: two-day window (0, 1), three-day window (-1, 1) and four-day window (-2, 1). These short time and around the announcement day windows are used to test for the liquidity assumption of the market where the acquirer is from. There are two other groups of windows used in this research *(described on the 4.3.1.1. section).* In general, these three groups show the similar patterns of findings but the group testing for the market's liquidity performs best with many statistically significant results. That is the reason for choosing this group in order to denote findings in this research. *(The other two groups results are available on request).*

The independent variables are control, time trend, listed target, non-diversified, post-acquisition ownership (x %), post-acquisition ownership (x >=95%) and transaction value. The time-trend variable is centered at the year 2005. This variable indicates the relative time of the individual transaction in the sample timeline. The M&A transactions in the year 1995, 2005 and 2015 are transferred to the value of -10, 0 and 10 for this time trend variable respectively. Listed target is a dummy variable in which the acquisition with listed target will receive the value of 1, otherwise 0. Non-diversified is a dummy variable denotes whether or not involvers are in the same three-digit SIC industry code. Post-acquisition ownership (x %) and transaction value are log-transformed. Post-acquisition ownership (x >=95%) is a dummy variable taking the value of 1 if the acquirer possesses from 95% control of target and 0 otherwise.

**Table 7: Majority control drives positive returns for developed-market acquirers**

| | Windows | (1) | (2) | (3) | (4) | (5) | (6) | (7) | (8) |
|---|---|---|---|---|---|---|---|---|---|
| Control | (0,1) | 0.020*** | 0.020*** | 0.020*** | 0.021*** | | 0.030*** | 0.026*** | 0.021** |
| | | (0.005) | (0.005) | (0.006) | (0.005) | | (0.010) | (0.006) | (0.008) |
| | (-1,1) | 0.027*** | 0.027*** | 0.027*** | 0.027*** | | 0.051*** | 0.038*** | 0.029*** |
| | | (0.008) | (0.008) | (0.008) | (0.008) | | (0.014) | (0.009) | (0.010) |
| | (-2,1) | 0.024*** | 0.024*** | 0.027*** | 0.024*** | | 0.044*** | 0.035*** | 0.026** |

|  |  | (0.008) | (0.008) | (0.008) | (0.008) |  | (0.015) | (0.010) | (0.010) |
|---|---|---|---|---|---|---|---|---|---|
| **Time-trend** | (0,1) |  | -0.000 |  |  |  |  |  |  |
|  |  |  | (0.001) |  |  |  |  |  |  |
|  | (-1,1) |  | 0.001 |  |  |  |  |  |  |
|  |  |  | (0.001) |  |  |  |  |  |  |
|  | (-2,1) |  | 0.001 |  |  |  |  |  |  |
|  |  |  | (0.001) |  |  |  |  |  |  |
| **Listed target** | (0,1) |  |  | -0.001 |  |  |  |  |  |
|  |  |  |  | (0.008) |  |  |  |  |  |
|  | (-1,1) |  |  | 0.002 |  |  |  |  |  |
|  |  |  |  | (0.012) |  |  |  |  |  |
|  | (-2,1) |  |  | 0.011 |  |  |  |  |  |
|  |  |  |  | (0.013) |  |  |  |  |  |
| **Non-Diversified** | (0,1) |  |  |  | -0.006 |  |  |  |  |
|  |  |  |  |  | (0.005) |  |  |  |  |
|  | (-1,1) |  |  |  | -0.003 |  |  |  |  |
|  |  |  |  |  | (0.008) |  |  |  |  |
|  | (-2,1) |  |  |  | 0.003 |  |  |  |  |
|  |  |  |  |  | (0.008) |  |  |  |  |
| **Post-acquisition ownership (x %)** | (0,1) |  |  |  |  |  | **0.017**\*\*\* | -0.013 |  |
|  |  |  |  |  |  |  | **(0.007)** | (0.012) |  |
|  | (-1,1) |  |  |  |  |  | **0.018**\* | **-0.034**\*\* |  |
|  |  |  |  |  |  |  | **(0.010)** | **(0.017)** |  |
|  | (-2,1) |  |  |  |  |  | **0.017**\* | -0.029 |  |
|  |  |  |  |  |  |  | **(0.010)** | (0.018) |  |
| **Post-acquisition Ownership (x >=95%)** | (0,1) |  |  |  |  |  |  | -0.011 |  |
|  |  |  |  |  |  |  |  | (0.008) |  |
|  | (-1,1) |  |  |  |  |  |  | **-0.023**\*\* |  |
|  |  |  |  |  |  |  |  | **(0.011)** |  |
|  | (-2,1) |  |  |  |  |  |  | -0.022 |  |
|  |  |  |  |  |  |  |  | (0.012) |  |
| **Transaction value** | (0,1) |  |  |  |  |  |  |  | 0.005 |
|  |  |  |  |  |  |  |  |  | (0.004) |
|  | (-1,1) |  |  |  |  |  |  |  | 0.005 |
|  |  |  |  |  |  |  |  |  | (0.005) |
|  | (-2,1) |  |  |  |  |  |  |  | 0.002 |
|  |  |  |  |  |  |  |  |  | (0.005) |
| **Adj. R-square** | (0,1) | 0.115 | 0.106 | 0.107 | 0.118 | 0.050 | 0.116 | 0.123 | 0.094 |
|  | (-1,1) | 0.094 | 0.090 | 0.086 | 0.087 | 0.022 | 0.118 | 0.122 | 0.093 |
|  | (-2,1) | 0.070 | 0.067 | 0.068 | 0.062 | 0.017 | 0.083 | 0.091 | 0.076 |
| **N** |  | 109 | 112 | 112 | 112 | 112 | 112 | 112 | 63 |

*This table reveals the results of regressions on CARs and other characteristics of acquirer in (0, 1), (-1, 1) and (-2, 1) windows around the announcement date. All M&A transactions in the sample were announced between 1995 and 2015 and involve a public acquirer from a developed country and a Vietnamese target (sample 1). Control is a dummy variable that takes the value of 1 if the acquirer holds at least 50% target's shares after the transaction, and 0 otherwise. Time-trend is a continuous variable basing on the year the deal was announced, which is relative to the sample timeline. The time-trend has the central point at the year 2005 so that 2005 is year 0. Non-Diversify is a dummy variable denoting whether or not involvers are in the same three-digit SIC industry code. Listed target is a dummy variable that takes the value of 1 if the target is listed in Vietnam's stock exchange, and 0 otherwise. Post-acquisition ownership and Transaction Value are log-transformed. Mean estimated coefficients are reported and standard errors are in parentheses. ∗,∗∗ and ∗∗∗ indicate statistically significant levels at 10%, 5%, and 1% respectively.*

Column 1 shows the results of the regressions between three measurements of acquirer's return and the control variable. The coefficients are 2.0%, 2.7% and 2.4% for the three windows (0, 1), (-1, 1) and (-2, 1) respectively with the significant level at 1% for all these coefficients. In column 2, we use the same regressions in column 1 but adding the time-trend variable in order to investigate whether the return of controlling Vietnamese target by developed-market acquirer varies over time. The results suggest that there is no significant evidence for the change of return over time.

The regressions in column 3 add the listed target variable to the regression in column 1 in order to investigate the effect of acquiring listed target. The results are not significant. This result is not consistent with the pattern of return for acquisition from developed-market to emerging-market in the study of Chari et al. (2010). This may reflect the different effect between taking-control and non-taking-control acquisition. The fact is that the acquisition involving listed target could not be a taking-control acquisition as the possession of acquirer in Vietnam listed target is restricted at lower than 50% share after the transaction.

Furthermore, column 4 shows the insignificant result for the non-diversify variable. The acquisition involving firms from the same or different industries do not make the different return. This result is consistent with the pattern of return for acquisition from developed-market to emerging-market in the study of Chari et al. (2010).

The proportion of share owned by the acquirer in the target after the M&A transaction may have a relationship with the control variable. The control is defined that the acquirer possesses at least 50% share of the target. In reality, considering some target in which the shareholder structure is diffuse, the acquirer could take the control with the significantly smaller proportion of share than 50%. Furthermore, the proportion of share owned by the acquirer after transaction may affect the return of acquirer. Holding 10% may have a lower effect than holding 80% of total shares. In order to ensure the robustness in measuring control, the regression with the alternative measuring of control is adopted.

The Post-acquisition ownership (x %) variable is used. This variable is a continuous measure of ownership and is log-transformed of the shares held by acquirer after the transaction. The coefficient for this variable is positive and significant and reported in column 5 of Table 7. The coefficients for the three windows (0, 1), (-1, 1) and (-2, 1) are 1.7%, 1.8% and 1.7 % with the significant level at 1%, 10% and 10% respectively.

Considering the taking-control transactions, the mean and median of the proportion of shares possessed by acquirer after the transaction are 81.88% and 90% respectively. In order to test for whether the positive return of acquirer is driven by the complete or nearly complete M&A transactions, the dummy variable named "dummy95" is added to the regression with dummy variable control. "Dummy95" will get the value of 1 if after the transaction the acquirer possesses from 95% of target's share, and 0 otherwise. The result in column 7 of Table 7 suggests the effect of control is still significantly positive, but the complete, or nearly complete possession of acquirer do not have a significant effect.

Finally, the column 8 of Table 7 shows the result of testing whether the control is simply a proxy of transaction value. The control has a positive and significant effect while there is no clear evidence that the transaction value drives the return. In summary, the results in column 5, 6, 7 and 8 are consistent with the hypothesis that the control in emerging-market target drives the positive return of developed-market acquirer but that control is not simply the proxy of the amount of share possessing after the transaction or the size of that transaction.

### 5.2. Dollar value gains for developed-market acquirers' shareholders

**Table 8: Summary statistics for value gains by acquirers**

|  | CAR (0,1) | CAR (-1,1) | CAR (-2,1) | Acquirer market capitalization ($M) | Dollar value gain per transaction ($M) | Transaction value ($M) | Net synergy return per transaction |
|---|---|---|---|---|---|---|---|
| Panel A: Developed-market acquirers gain majority control | | | | | | | |
| Mean | 1.330% | 2.061% | 2.026% | 9760.107 | 44.953 | 47.837 | -2.486 |
| Median | 0.670% | 1.025% | 1.150% | 1943.755 | 2.685 | 23.730 | 0.212 |
| Top quartile | 2.235% | 3.210% | 2.985% | 5531.593 | 47.477 | 56.473 | 3.065 |
| Bottom quartile | -0.448% | -0.338% | -0.248% | 450.743 | -8.255 | 2.783 | -0.330 |
| Std dev | 2.967% | 4.722% | 5.027% | 22811.01 | 175.465 | 76.905 | 22.041 |
| N | 50 | 50 | 50 | 50 | 50 | 26 | 26 |
| Panel B: Developed-market acquirers do not gain majority control | | | | | | | |
| Mean | -0.720% | -0.611% | -0.401% | 29144.059 | -29.241 | 35.084 | -8.204 |
| Median | -0.150% | -0.180% | -0.315% | 8900.475 | -2.880 | 17.325 | -0.038 |
| Top quartile | 0.483% | 0.950% | 1.158% | 31427.150 | 83.952 | 49.559 | 2.405 |
| Bottom quartile | -1.633% | -1.655% | -1.205% | 1737.163 | -136.930 | 2.344 | -7.429 |
| Std dev | 2.566% | 3.244% | 3.339% | 43229.628 | 752.026 | 47.567 | 50.853 |
| N | 62 | 62 | 62 | 62 | 62 | 37 | 37 |

*This table shows the acquirer's shareholder value gains from the announcement of M&A deals involving three sample (DM-VN, EE-VN, VN-VN). CAR is based on the (0, 1), (-1, 1) and (-2, 1) windows around the deal announcement date. Dollar value gains are calculated by multiplying the acquirer ARs in three days by their one-day-before market capitalization and then summing these three value. Net synergy returns are calculated by dividing the dollar value gain by the transaction value.*

Table 8 gives the distribution of the magnitude of the developed-market acquirer's shareholder value gaining from the announcement of M&A deals to Vietnamese target (DM-VN sample). While panel A presents the value gaining by acquirers from M&A transactions in which the acquirer gains the control, panel B presents the value gaining by acquirers from M&A transactions in which the acquirer does not gain the control.

Panel A shows medians of CARs for (0, 1), (-1, 1) and (-2, 1) windows are positive and equal 0.67%, 1.025% and 1.150% respectively in transactions where the control is acquired. In contrast, in Panel B, the medians of CARs for (0, 1), (-1, 1) and (-2, 1) windows are negative and equal -0.150%, -0.180% and -0.315% respectively in transactions where the control is not acquired. A Wilcoxon signed-rank test of medians shows that transactions, where the acquirer gains the majority control, offers a significantly higher return compared to the transactions, where the control is not acquired.

These numbers suggest that the shareholders of developed-market acquirers reap significant dollar value gains from the transactions to Vietnam when the control is acquired. Considering all acquisitions transferring the control to developed-market acquirers and Vietnamese targets in our sample, the returns from announcements translate into an aggregate dollar value gain is about $2.248 billion for the shareholders of developed-market acquirers.

In addition, the median of dollar value gain per deal, where the acquirer gains the majority control, is $2.685 million comparing with $23.730 million for transaction value of these deals. The median for net synergy return per transaction (calculated by acquirer's dollar value gain/transaction value) is 0.212. In other words, in median, the stock market expects that developed-market acquirers will gain 21 cents at present value for every dollar they spend on taking control in emerging-market acquisitions.

By contrast, the median of net synergy return per non-taking-control transaction is -0.038 (Table 8, panel B). This number suggests that the key mechanism for generating positive returns for developed-market investors in Vietnamese market is a transfer of control.

The huge value gained by developed-market acquirers when acquiring firms in Vietnam address a question for the source from which these values are. While this phenomenon is not consistent with many previous researches in domestic M&A transactions in developed market (e.g. Moeller et al.,2004), with the different approach, by focusing on cross-border deal to emerging market, Chari et al. (2010) find evidence on the substantial value gain shipping from emerging market to developed one by taking control the target.

### 5.3. Are the gains associated with transferring the control of Vietnamese target unique to developed-market acquirers?

The results above suggest that the key mechanism for generating positive returns for developed-market investors in Vietnamese market is a transfer of control. In Table 9, whether this positive result is unique to the developed-market acquirers or there is an alternative explanation for the finding will be explored in detailed.

In column 1 of Table 9, all the three samples are combined. The acquirer is from either developed market or emerging market including Vietnam. The regression is run on the number of variables including: control, DM acquirer, interactive term between control and DM acquirer (Control*DM acquirer), publicly listed, non-diversified, market capitalization (MV), interactive firm between control and market capitalization (Control*MV). The coefficients of control are positive, but they are not significant for the three windows. The acquirer taking the control in Vietnamese target seems to get higher return compared with the non-taking-control one. However, the coefficients for interactive term between control and DM acquirer are positive and significant with the value of 2.6%, 3.3% and 2.6% for the three windows (0,1), (-1,1) and (-2,1) respectively. These three values are statistically significant at the 1%, 5%, and 10% respectively. These results is the first support for the argument that the gains associated with transferring the control of Vietnamese target is unique to developed-market acquirers

Column 2 includes the acquirer from the sample 1 (DM-VN) and 2 (EM-VN) denoting the cross-border deal only. Running the regression similar to the one in column 1, the results are also similar.

Column 3 includes the acquirers from Vietnam denoting the domestic deals only. Running the regression similar to the one in column 1 but excluding DM acquirer and Control*DM acquirer, the result suggests that the control of Vietnamese acquirer to Vietnamese target does not have the significant meaning.

Reviewing the table 4 it is worth noting that the acquirer from the developed market has the median for market capitalization is about 10 times higher than one from emerging market (excluding Vietnam) and 80 times higher than Vietnam acquirer. The median market capitalization for developed-market, emerging-market and Vietnam acquirer are $3764.355 million, $329.69 million, $40.255 million respectively. There is evidence for acquirer's size effect in either domestic (e.g. Moeller et al.,2004) or cross-border M&A transaction (e.g. Chari 2010). In the form of log-transferring (MV), and interactive term between control and market capitalization (MV*Control), the coefficient for these terms are negative (Table 10, column 1, 2 and 3). This result is consistent with literature in either the domestic or cross-border transaction. However, the result is not statistically significant.

In summary, considering the M&A transactions in Vietnam, the positive return for acquirer is unique for developed-market acquirers when they gain the control of the target. Either the cross-border deals from emerging-countries or the domestic transactions do not have the clear benefit.

**Table 9: Gains from majority control are unique to developed-market acquirers**

|  | Windows | 1 (All-VN) | 2 (CB-VN) | 3 (VN-VN) |
|---|---|---|---|---|
| Sample description |  | All acquirers and Vietnam targets | Foreign acquirers and Vietnam targets | Vietnam acquirers and Vietnam targets |
| **Control** | (0,1) | 0.007 | 0.017 | 0.003 |
|  |  | (0.008) | (0.018) | (0.012) |
|  | (-1,1) | 0.015 | 0.020 | -0.000 |
|  |  | (0.011) | (0.028) | (0.015) |
|  | (-2,1) | 0.015 | -0.011 | 0.012 |
|  |  | (0.012) | (0.030) | (0.017) |
| **DM Acquirer** | (0,1) | -0.009 | -0.008 |  |
|  |  | (0.008) | (0.011) |  |
|  | (-1,1) | 0.001 | -0.009 |  |
|  |  | (0.010) | (0.017) |  |
|  | (-2,1) | 0.005 | -0.014 |  |
|  |  | (0.012) | (0.018) |  |
| **Control*DM Acquirer** | (0,1) | **0.026\*\*** | **0.026\*** |  |
|  |  | **(0.010)** | **(0.014)** |  |
|  | (-1,1) | **0.033\*\*** | **0.051\*\*** |  |
|  |  | **(0.014)** | **(0.022)** |  |
|  | (-2,1) | **0.026\*** | **0.053\*\*** |  |
|  |  | **(0.015)** | **(0.023)** |  |
| **Publicly listed** | (0,1) | 0.002 | 0.000 | 0.002 |
|  |  | (0.04) | (0.008) | (0.006) |
|  | (-1,1) | 0.003 | 0.002 | 0.003 |
|  |  | (0.006) | (0.013) | (0.007) |
|  | (-2,1) | 0.005 | 0.012 | 0.003 |
|  |  | (0.007) | (0.013) | (0.008) |
| **Non-Diversified** | (0,1) | -0.005 | -0.005 | -0.003 |
|  |  | (0.004) | (0.005) | (0.006) |
|  | (-1,1) | -0.004 | -0.003 | -0.003 |
|  |  | (0.005) | (0.008) | (0.008) |
|  | (-2,1) | -0.001 | 0.003 | -0.006 |
|  |  | (0.006) | (0.008) | (0.009) |
| **MV** | (0,1) | 0.003 | 0.005 | 0.001 |
|  |  | (0.003) | (0.003) | (0.004) |
|  | (-1,1) | 0.001 | 0.004 | -0.003 |
|  |  | (0.004) | (0.005) | (0.005) |
|  | (-2,1) | 0.000 | -0.001 | -0.002 |
|  |  | (0.004) | (0.006) | (0.006) |
| **MV*Control** | (0,1) | -0.003 | -0.006 | 0.001 |
|  |  | (0.004) | (0.005) | (0.006) |
|  | (-1,1) | -0.006 | -0.013 | 0.004 |
|  |  | (0.005) | (0.007) | (0.008) |
|        | (-2,1) | -0.005 | -0.005 | -0.001 |
|        |        | (0.006) | (0.008) | (0.009) |
| **Adj. R sqt** | (0,1) | 0.027 | 0.091 | -0.026 |
|        | (-1,1) | 0.041 | 0.081 | -0.022 |
|        | (-2,1) | 0.024 | 0.043 | -0.017 |
| **N**  |        | 308   | 134   | 177   |

*This table shows the results of regressions where the dependent variables are CARs for acquirer firms during (0, 1), (-1, 1) and (-2, 1) windows around the announcement date on involved firms' characteristics. Control is a dummy variable that take the value of 1 if the acquirer holds at least 50% target's shares after the transaction, and 0 otherwise. DM acquirer is a dummy variable that take the value of 1 if the acquirer is from the developed-market, and 0 otherwise. Control\*DM acquirer is an interactive term between 2 variables Control and DM acquirer. Listed target is a dummy variable that takes the value of 1 if the target is listed in Vietnam stock exchange, and 0 otherwise. Non-Diversified deal is a dummy variable denotes whether or not involved firms have the same three-digit SIC industry code. MV is acquirer market capitalization, which is measured on the day of acquisition announcement, is log-transformed. MV\*control is an interactive term between 2 variables MV and Control. ∗, ∗∗ and ∗∗∗ indicate statistically significant levels at the 10%, 5%, and 1% respectively.*

## 6. Conclusion

This paper gives a glance at the M&A activities in Vietnam from 1995 to 2015. By reviewing some historical events in economic, political, and social approaches, the transformation from a country with backward economy to an emerging market is disclosed. This paper uses the event study with OLS market model approach combining with multivariate regression in order to identify the abnormal return of M&A transactions in Vietnam. The unique benefit of developed-market acquirer when taking control Vietnam target is an interesting finding. Moreover, the estimation of the dollar value gained by developed-market acquirers shows the benefit from M&A deals and also strengthen the expected findings of this research.

Vietnam is an emerging country so it shares some characteristics with other emerging countries. In general, we could expect that doing M&A in Vietnam is similar to that in any other emerging countries. The abnormal returns of acquirers from developed countries are compensation for their improvements in target's business in order to overcome difficulties in its market (Francis et al., 2008) or weak target's corporate governance (Chari et al., 2010).

However, everything has its limitation. Although many potential errors are excluded by careful and cautions consideration in this research, are still small amount of issues. These issues will be a prospect for future research. The unavailable data for M&A taking-control transactions between developed-market acquirers and the listed targets could lead to a lack of evidence of value gains in targets. The reason is that Vietnam's law on foreign investment did not allow the foreign investors including individuals, institutions, and funds, and so on hold more than 50% of Vietnam's public firm before September 2015 (there is a foreign holding cap at 50%). From September 2015 onward, this cap was removed and the foreign acquirers could hold up to 100 percent (see Vietnam, The Government, 2015). However, this removal needs time to be completely effective. Hopefully, the significant number of M&A transactions taking control listed Vietnam firms by developed-market acquirers will happen in the near future so that the gains of Vietnam target can be confirmed.

Moreover, there is a limitation in choosing event study method as mentioned in literature review part. Regarding the efficient market theory (see Fama, 1970), the market reflects the fundamental factors efficiently just when the market is in strong form. The truth is that not all the stock markets in the world are in strong form so that the biased use of this method cannot provide the sufficient evaluation. The change in stock price also depends on market's expectation which needs time to be proven. An alternative to this method is to use accounting-based analysis, but as mentioned before, there are some issues in this alternative method requiring further time and effort to be adopted. These limitations lead us to conduct possibly further research in order to fully capturing the M&A transactions, especially a full picture in Vietnam.

For the future research, with the fast development of new technologies, especially in machine learning and artificial intelligence (e.g. Wolohan et al., 2018; Jurgens et al., 2019, Van et al., 2019, 2020), we can also gather information from social media to evaluate the effect of M&As to acquirers. For example, there will be a lot of arguments and comments around M&A announcements. Especially in social media platforms (e.g. Twitter). Perhaps the investors, and other market participants could predict the value from the acquisitions. This could offer potential future research to strengthen the results from this work.

## 8. Appendix

### Table 10: Sample description.

| | ALL<br>N=308 | DM-VN<br>N=112 | EM-VN<br>N=22 | VN-VN<br>N=174 |
|---|---|---|---|---|
| (0,1) | 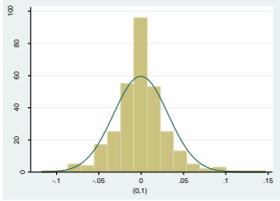<br>S=0.465, K=6.098 | 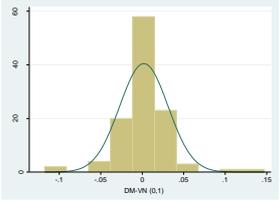<br>S=0.475, K=10.701 | 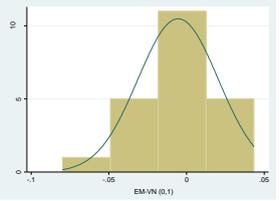<br>S=−0.902, K=4.553 | 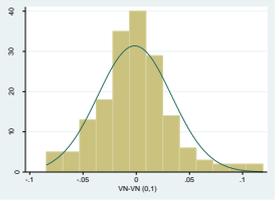<br>S=0.538, K=4.388 |
| (−1,1) | 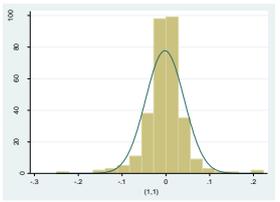<br>S=−0.016, K=10.342 | 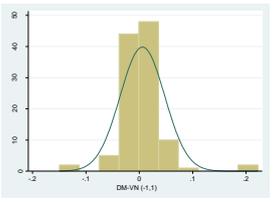<br>S=1.609, K=14.359 | 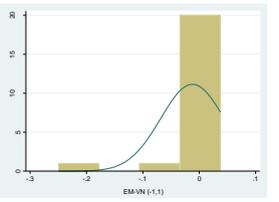<br>S=−3.551, K=15.625 | 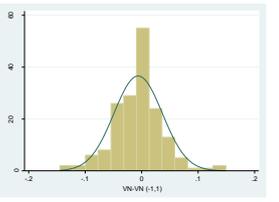<br>S=0.618, K=4.599 |
| (−2,1) | 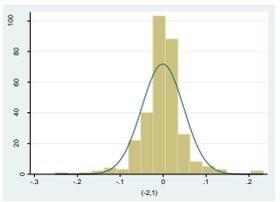<br>S=0.039, K=8.580 | 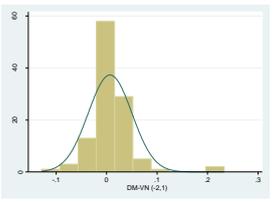<br>S=2.097, K=14.338 | 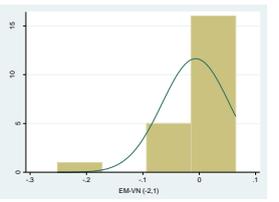<br>S=−3.257, K=14.403 | 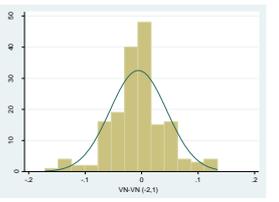<br>S=−0.058, K=4.203 |

N= number of observations, S=Skewness, K=Kurtosis.

**Table 11: Country classification**

| Data for calendar year: | 1995 | 1996 | 1997 | 1998 | 1999 | 2000 | 2001 | 2002 | 2003 | 2004 | 2005 | 2006 | 2007 | 2008 | 2009 | 2010 | 2011 | 2012 | 2013 | 2014 | 2015 |
|---|---|---|---|---|---|---|---|---|---|---|---|---|---|---|---|---|---|---|---|---|---|
| Australia | H | H | H | H | H | H | H | H | H | H | H | H | H | H | H | H | H | H | H | H | H |
| Canada | H | H | H | H | H | H | H | H | H | H | H | H | H | H | H | H | H | H | H | H | H |
| Denmark | H | H | H | H | H | H | H | H | H | H | H | H | H | H | H | H | H | H | H | H | H |
| France | H | H | H | H | H | H | H | H | H | H | H | H | H | H | H | H | H | H | H | H | H |
| Germany | H | H | H | H | H | H | H | H | H | H | H | H | H | H | H | H | H | H | H | H | H |
| Hong Kong SAR, China | H | H | H | H | H | H | H | H | H | H | H | H | H | H | H | H | H | H | H | H | H |
| Japan | H | H | H | H | H | H | H | H | H | H | H | H | H | H | H | H | H | H | H | H | H |
| Korea, Rep. | H | H | H | UM | UM | UM | H | H | H | H | H | H | H | H | H | H | H | H | H | H | H |
| Singapore | H | H | H | H | H | H | H | H | H | H | H | H | H | H | H | H | H | H | H | H | H |
| Sweden | H | H | H | H | H | H | H | H | H | H | H | H | H | H | H | H | H | H | H | H | H |
| Switzerland | H | H | H | H | H | H | H | H | H | H | H | H | H | H | H | H | H | H | H | H | H |
| Taiwan, China | H | H | H | H | H | H | H | H | H | H | H | H | H | H | H | H | H | H | H | H | H |
| United Kingdom | H | H | H | H | H | H | H | H | H | H | H | H | H | H | H | H | H | H | H | H | H |
| United States | H | H | H | H | H | H | H | H | H | H | H | H | H | H | H | H | H | H | H | H | H |
| India | L | L | L | L | L | L | L | L | L | L | L | L | LM | LM | LM | LM | LM | LM | LM | LM | LM |
| Indonesia | LM | LM | LM | LM | LM | LM | LM | LM | LM | LM | LM | LM | LM | LM | LM | LM | LM | LM | LM | LM | LM |
| Malaysia | UM | UM | UM | UM | UM | UM | UM | UM | UM | UM | UM | UM | UM | UM | UM | UM | UM | UM | UM | UM | UM |
| Myanmar | L | L | L | L | L | L | L | L | L | L | L | L | L | L | L | L | L | L | L | L | LM |
| Philippines | LM | LM | LM | LM | LM | LM | LM | LM | LM | LM | LM | LM | LM | LM | LM | LM | LM | LM | LM | LM | LM |
| Russian Federation | LM | LM | LM | LM | LM | LM | LM | LM | LM | LM | LM | UM | UM | UM | UM | UM | UM | UM | UM | UM | UM |
| Thailand | LM | LM | LM | LM | LM | LM | LM | LM | LM | LM | LM | LM | LM | LM | LM | LM | LM | UM | UM | UM | UM |
| Vietnam | L | L | L | L | L | L | L | L | L | L | L | L | L | L | LM | LM | LM | LM | LM | LM | LM |

*Source: World Bank (2016)*

**Table 12: Thresholds for country classification**

| Data for calendar year : | 1995 | 1996 | 1997 | 1998 | 1999 | 2000 | 2001 |
|---|---|---|---|---|---|---|---|
| Low income (L) | <= 765 | <= 785 | <= 785 | <= 760 | <= 755 | <= 755 | <= 745 |
| Lower middle income (LM) | 766-3,035 | 786-3,115 | 786-3,125 | 761-3,030 | 756-2,995 | 756-2,995 | 746-2,975 |
| Upper middle income (UM) | 3,036-9,385 | 3,116-9,645 | 3,126-9,655 | 3,031-9,360 | 2,996-9,265 | 2,996-9,265 | 2,976-9,205 |
| High income (H) | > 9,385 | > 9,645 | > 9,655 | > 9,360 | > 9,265 | > 9,265 | > 9,205 |
| Data for calendar year : | 2002 | 2003 | 2004 | 2005 | 2006 | 2007 | 2008 |
| Low income (L) | <= 735 | <= 765 | <= 825 | <= 875 | <= 905 | <= 935 | <= 975 |
| Lower middle income (LM) | 736-2,935 | 766-3,035 | 826-3,255 | 876-3,465 | 906-3,595 | 936-3,705 | 976-3,855 |
| Upper middle income (UM) | 2,936-9,075 | 3,036-9,385 | 3,256-10,065 | 3,466-10,725 | 3,596-11,115 | 3,706-11,455 | 3,856-11,905 |
| High income (H) | > 9,075 | > 9,385 | > 10,065 | > 10,725 | > 11,115 | > 11,455 | > 11,905 |
| Data for calendar year : | 2009 | 2010 | 2011 | 2012 | 2013 | 2014 | 2015 |
| Low income (L) | <= 995 | <= 1,005 | <= 1,025 | <= 1,035 | <= 1,045 | <= 1,045 | <= 1,025 |
| Lower middle income (LM) | 996-3,945 | 1,006-3,975 | 1,026-4,035 | 1,036-4,085 | 1,046-4,125 | 1,046-4,125 | 1,026-4,035 |
| Upper middle income (UM) | 3,946-12,195 | 3,976-12,275 | 4,036-12,475 | 4,086-12,615 | 4,126-12,745 | 4,126-12,735 | 4,036-12,475 |
| High income (H) | > 12,195 | > 12,275 | > 12,475 | > 12,615 | > 12,745 | > 12,735 | > 12,475 |

The numbers define GNI per capita in US$ (Atlas methodology)

*Source: World Bank (2016)*